%% file: main.tex
\pdfoutput=1

\documentclass[12pt,a4paper]{article}

\usepackage{ifthen} 
\newboolean{pdflatex}
\setboolean{pdflatex}{true} 

\newboolean{articletitles}
\setboolean{articletitles}{true} 

\newboolean{uprightparticles}
\setboolean{uprightparticles}{false} 

\newboolean{inbibliography}
\setboolean{inbibliography}{false} 
\usepackage{placeins}

\input{preamble}
\usepackage{longtable} 

\begin{document}

\renewcommand{\thefootnote}{\fnsymbol{footnote}}
\setcounter{footnote}{1}

\input{title-LHCb-PAPER}


\renewcommand{\thefootnote}{\arabic{footnote}}
\setcounter{footnote}{0}



\pagestyle{plain} 
\setcounter{page}{1}
\pagenumbering{arabic}


%

\input{introduction}

\input{detector}

\input{strategy}

\input{systematics}

\input{results}

\input{conclusion}

\input{acknowledgements}

\addcontentsline{toc}{section}{References}
\setboolean{inbibliography}{true}
\bibliographystyle{LHCb}
\bibliography{main,LHCb-PAPER,LHCb-CONF,LHCb-DP,LHCb-TDR}

\newpage
\input{LHCb_authorlist.tex}

\end{document}

%% file: preamble.tex
\usepackage{microtype}
\usepackage{lineno}  
\usepackage{xspace} 
\usepackage{caption} 

\usepackage{graphicx}  
\usepackage{color}
\usepackage{colortbl}
\graphicspath{{./figs/}} 

\usepackage{amsmath} 
\usepackage{amssymb}
\usepackage{amsfonts}
\usepackage{upgreek} 

\newcommand*\patchAmsMathEnvironmentForLineno[1]{%
\expandafter\let\csname old#1\expandafter\endcsname\csname #1\endcsname
\expandafter\let\csname oldend#1\expandafter\endcsname\csname
end#1\endcsname
 \renewenvironment{#1}%
   {\linenomath\csname old#1\endcsname}%
   {\csname oldend#1\endcsname\endlinenomath}%
}
\newcommand*\patchBothAmsMathEnvironmentsForLineno[1]{%
  \patchAmsMathEnvironmentForLineno{#1}%
  \patchAmsMathEnvironmentForLineno{#1*}%
}
\AtBeginDocument{%
\patchBothAmsMathEnvironmentsForLineno{equation}%
\patchBothAmsMathEnvironmentsForLineno{align}%
\patchBothAmsMathEnvironmentsForLineno{flalign}%
\patchBothAmsMathEnvironmentsForLineno{alignat}%
\patchBothAmsMathEnvironmentsForLineno{gather}%
\patchBothAmsMathEnvironmentsForLineno{multline}%
\patchBothAmsMathEnvironmentsForLineno{eqnarray}%
}

\usepackage{hyperref}    
\usepackage[all]{hypcap} 

\input{lhcb-symbols-def} 

\usepackage{cite} 
\usepackage{mciteplus}

%% file: lhcb-symbols-def.tex
\def\TopoM      {\ensuremath{M}\xspace}                 
\def\TopoMCorr  {\ensuremath{M_{\text{corr}}}\xspace}                 
\def\Pmiss      {\ensuremath{p\sin{\theta}}\xspace}
\def\Pmisssq    {\ensuremath{p^2\sin^2{\theta}}\xspace}
\def\antikt     {\ensuremath{\text{anti-}k_{\mathrm{T}}}\xspace}                 

\def\lhcb {\mbox{LHCb}\xspace}
\def\ux85 {\mbox{UX85}\xspace}

\ifthenelse{\boolean{uprightparticles}}%
{
 
 \def\Pgamma      {\ensuremath{\upgamma}\xspace}

 \def\Peta        {\ensuremath{\upeta}\xspace}

 \def\Pmu         {\ensuremath{\upmu}\xspace}

 \def\Ppi         {\ensuremath{\uppi}\xspace}

 \def\Pphi        {\ensuremath{\upphi}\xspace}

 \def\Ppsi        {\ensuremath{\uppsi}\xspace}

 \def\PDelta      {\ensuremath{\Delta}\xspace}                 
 \def\PXi      {\ensuremath{\Xi}\xspace}                 
 \def\PLambda      {\ensuremath{\Lambda}\xspace}                 
 \def\PSigma      {\ensuremath{\Sigma}\xspace}                 
 \def\POmega      {\ensuremath{\Omega}\xspace}                 
 \def\PUpsilon      {\ensuremath{\Upsilon}\xspace}                 
 
 \def\PB      {\ensuremath{\mathrm{B}}\xspace}                 
                  
 \def\PD      {\ensuremath{\mathrm{D}}\xspace}

 \def\PJ      {\ensuremath{\mathrm{J}}\xspace}                 
 \def\PK      {\ensuremath{\mathrm{K}}\xspace}

 \def\Pb      {\ensuremath{\mathrm{b}}\xspace}                 
 \def\Pc      {\ensuremath{\mathrm{c}}\xspace}

 \def\Pi      {\ensuremath{\mathrm{i}}\xspace}

}
{
 
 \def\Pgamma      {\ensuremath{\gamma}\xspace}

 \def\Peta        {\ensuremath{\eta}\xspace}

 \def\Pmu         {\ensuremath{\mu}\xspace}

 \def\Ppi         {\ensuremath{\pi}\xspace}

 \def\Pphi        {\ensuremath{\phi}\xspace}

 \def\Ppsi        {\ensuremath{\psi}\xspace}                 
                  
 \mathchardef\PDelta="7101
 \mathchardef\PXi="7104
 \mathchardef\PLambda="7103
 \mathchardef\PSigma="7106
 \mathchardef\POmega="710A
 \mathchardef\PUpsilon="7107
                  
 \def\PB      {\ensuremath{B}\xspace}                 
                  
 \def\PD      {\ensuremath{D}\xspace}

 \def\PJ      {\ensuremath{J}\xspace}                 
 \def\PK      {\ensuremath{K}\xspace}

 \def\Pb      {\ensuremath{b}\xspace}                 
 \def\Pc      {\ensuremath{c}\xspace}

 \def\Pi      {\ensuremath{i}\xspace}

}

\def\mup        {\ensuremath{\Pmu^+}\xspace}
\def\mun        {\ensuremath{\Pmu^-}\xspace} 
\def\mumu       {\ensuremath{\Pmu^+\Pmu^-}\xspace}

\def\g      {\ensuremath{\Pgamma}\xspace}

\def\cquark    {\ensuremath{\Pc}\xspace}

\def\bquark    {\ensuremath{\Pb}\xspace}

\def\pion  {\ensuremath{\Ppi}\xspace}

\def\pipm  {\ensuremath{\pion^\pm}\xspace}

\def\kaon  {\ensuremath{\PK}\xspace}
  \def\Kbar  {\kern 0.2em\overline{\kern -0.2em \PK}{}\xspace}

\def\Kz    {\ensuremath{\kaon^0}\xspace}
\def\Kzb   {\ensuremath{\Kbar^0}\xspace}
\def\KzKzb {\ensuremath{\Kz \kern -0.16em \Kzb}\xspace}
\def\Kp    {\ensuremath{\kaon^+}\xspace}
\def\Km    {\ensuremath{\kaon^-}\xspace}
\def\Kpm   {\ensuremath{\kaon^\pm}\xspace}
\def\Kmp   {\ensuremath{\kaon^\mp}\xspace}
\def\KpKm  {\ensuremath{\Kp \kern -0.16em \Km}\xspace}
\def\KS    {\ensuremath{\kaon^0_{\rm\scriptscriptstyle S}}\xspace}

  \def\Dbar    {\kern 0.2em\overline{\kern -0.2em \PD}{}\xspace}
\def\D       {\ensuremath{\PD}\xspace}

\def\Dz      {\ensuremath{\D^0}\xspace}
\def\Dzb     {\ensuremath{\Dbar^0}\xspace}
\def\DzDzb   {\ensuremath{\Dz {\kern -0.16em \Dzb}}\xspace}
\def\Dp      {\ensuremath{\D^+}\xspace}
\def\Dm      {\ensuremath{\D^-}\xspace}
\def\Dpm     {\ensuremath{\D^\pm}\xspace}

\def\DpDm    {\ensuremath{\Dp {\kern -0.16em \Dm}}\xspace}

\def\B       {\ensuremath{\PB}\xspace}
\def\Bbar    {\ensuremath{\kern 0.18em\overline{\kern -0.18em \PB}{}}\xspace}

\def\Bpm     {\ensuremath{\B^\pm}\xspace}

\def\jpsi     {\ensuremath{{\PJ\mskip -3mu/\mskip -2mu\Ppsi\mskip 2mu}}\xspace}

  \def\Y#1S{\ensuremath{\PUpsilon{(#1S)}}\xspace}

\def\Lbar {\ensuremath{\kern 0.1em\overline{\kern -0.1em\PLambda}}\xspace}

\newcommand{\decay}[2]{\ensuremath{#1\!\to #2}\xspace}         

\def\to                 {\ensuremath{\rightarrow}\xspace}

\def\AT#1     {\ensuremath{A_{\mathrm{T}}^{#1}}\xspace}           

\def\C#1      {\ensuremath{\mathcal{C}_{#1}}\xspace}                       
\def\Cp#1     {\ensuremath{\mathcal{C}_{#1}^{'}}\xspace}                    
\def\Ceff#1   {\ensuremath{\mathcal{C}_{#1}^{\mathrm{(eff)}}}\xspace}        
\def\Cpeff#1  {\ensuremath{\mathcal{C}_{#1}^{'\mathrm{(eff)}}}\xspace}       
\def\Ope#1    {\ensuremath{\mathcal{O}_{#1}}\xspace}                       
\def\Opep#1   {\ensuremath{\mathcal{O}_{#1}^{'}}\xspace}                    

\newcommand{\tev}{\ensuremath{\mathrm{\,Te\kern -0.1em V}}\xspace}
\newcommand{\gev}{\ensuremath{\mathrm{\,Ge\kern -0.1em V}}\xspace}
\newcommand{\mev}{\ensuremath{\mathrm{\,Me\kern -0.1em V}}\xspace}
\newcommand{\kev}{\ensuremath{\mathrm{\,ke\kern -0.1em V}}\xspace}
\newcommand{\ev}{\ensuremath{\mathrm{\,e\kern -0.1em V}}\xspace}
\newcommand{\gevc}{\ensuremath{{\mathrm{\,Ge\kern -0.1em V\!/}c}}\xspace}
\newcommand{\mevc}{\ensuremath{{\mathrm{\,Me\kern -0.1em V\!/}c}}\xspace}
\newcommand{\gevcc}{\ensuremath{{\mathrm{\,Ge\kern -0.1em V\!/}c^2}}\xspace}
\newcommand{\gevgevcccc}{\ensuremath{{\mathrm{\,Ge\kern -0.1em V^2\!/}c^4}}\xspace}
\newcommand{\mevcc}{\ensuremath{{\mathrm{\,Me\kern -0.1em V\!/}c^2}}\xspace}

\def\mum  {\ensuremath{\,\upmu\rm m}\xspace}

\def\fb   {\ensuremath{\mbox{\,fb}}\xspace}
\def\invfb   {\ensuremath{\mbox{\,fb}^{-1}}\xspace}

\newcommand{\chisq}{\ensuremath{\chi^2}\xspace}

\def\gsim{{~\raise.15em\hbox{$>$}\kern-.85em
          \lower.35em\hbox{$\sim$}~}\xspace}
\def\lsim{{~\raise.15em\hbox{$<$}\kern-.85em
          \lower.35em\hbox{$\sim$}~}\xspace}

\def\PDF {PDF\xspace}

\def\sPlot{\mbox{\em sPlot}}

\def\sqs   {\ensuremath{\protect\sqrt{s}}\xspace}

\def\ptot       {\mbox{$p$}\xspace}
\def\pt         {\mbox{$p_{\rm T}$}\xspace}

\def\evtgen     {\mbox{\textsc{EvtGen}}\xspace}
\def\pythia     {\mbox{\textsc{Pythia}}\xspace}
\def\mcfm       {\mbox{\textsc{MCFM}}\xspace}
\def\fastjet    {\mbox{\textsc{FastJet}}\xspace}

\def\geant      {\mbox{\textsc{Geant4}}\xspace}

\def\tell1  {TELL1\xspace}
\def\ukl1   {UKL1\xspace}



%% file: title-LHCb-PAPER.tex

\begin{titlepage}
\pagenumbering{roman}

\vspace*{-1.5cm}
\centerline{\large EUROPEAN ORGANIZATION FOR NUCLEAR RESEARCH (CERN)}
\vspace*{1.5cm}
\hspace*{-0.5cm}
\begin{tabular*}{\linewidth}{lc@{\extracolsep{\fill}}r}
\ifthenelse{\boolean{pdflatex}}
{\vspace*{-2.7cm}\mbox{\!\!\!\includegraphics[width=.14\textwidth]{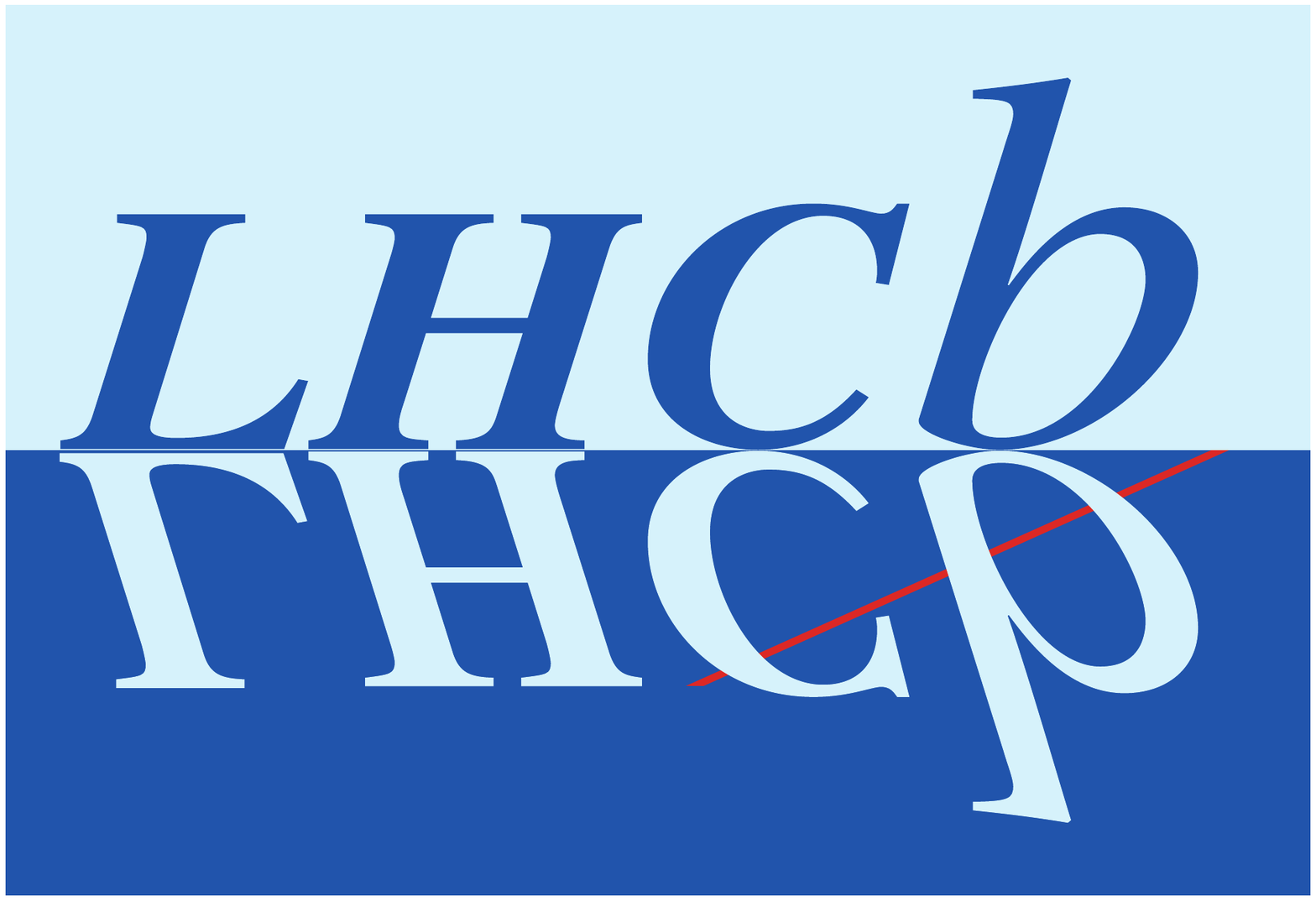}} & &}%
{\vspace*{-1.2cm}\mbox{\!\!\!\includegraphics[width=.12\textwidth]{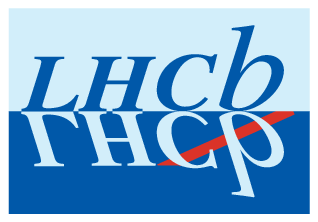}} & &}%
\\
 & & CERN-PH-EP-2014-267 \\  
 & & LHCb-PAPER-2014-055 \\  
 & & December 23, 2014 \\
 & & \\
\end{tabular*}

\vspace*{2.0cm}

{\bf\boldmath\huge
\begin{center}

Measurement of the Z+b-jet cross-section in pp collisions at $\sqrt{s}=7\tev$ in the forward region 
\end{center}
}

\vspace*{1.0cm}

\begin{center}
The LHCb collaboration\footnote{Authors are listed at the end of this paper.}
\end{center}

\vspace{\fill}

\begin{abstract}
  \noindent
  The associated production of a Z boson or an off-shell photon $\gamma^*$ with a bottom quark in the forward region is 
  studied using proton-proton collisions at a centre-of-mass energy of $7\tev$. 
  The Z bosons are reconstructed in the
  \decay{\text{Z}/\g^*}{\mup\mun} final state from muons with a transverse momentum larger than 20\gev, while two transverse 
  momentum thresholds are considered for jets (10\gev and 20\gev). 
  Both muons and jets are reconstructed in the pseudorapidity range $2.0 < \eta < 4.5$. 
  The results are based on data corresponding to 1.0\invfb
  recorded in 2011 with the LHCb detector.
  The measurement of the Z+b-jet cross-section is normalized to the Z+jet cross-section.   
  The measured cross-sections are
  \begin{equation*}
  \sigma(\text{$\text{Z}/\g^*(\mup\mun)$+b-jet}) = 295 \pm 60~(\text{stat}) \pm 51~(\text{syst}) \pm 10~(\text{lumi}) \fb
  \end{equation*}
  for $\pt$(jet)$>10\gev$, and
  \begin{equation*}
  \sigma(\text{$\text{Z}/\g^*(\mup\mun)$+b-jet}) =  128 \pm 36~(\text{stat}) \pm 22~(\text{syst}) \pm 5~(\text{lumi}) \fb
  \end{equation*}
  for $\pt$(jet)$>20\gev$.

\end{abstract}

\vspace*{1.0cm}

\begin{center}
  Published in JHEP 01 (2015) 064
\end{center}

\vspace{\fill}

{\footnotesize 
\centerline{\copyright~CERN on behalf of the \lhcb collaboration, license \href{http://creativecommons.org/licenses/by/4.0/}{CC-BY-4.0}.}}
\vspace*{2mm}

\end{titlepage}


\newpage
\setcounter{page}{2}
\mbox{~}


%
%

\cleardoublepage

%% file: introduction.tex

\section{Introduction}
\label{sec:Introduction}

The cross-section for the forward production of a Z boson\footnote{Throughout this paper Z boson includes both the Z$^0$ and the off-shell photon, $\gamma^*$, contributions.}
in association with a bottom quark (referred to as Z+b-jet) 
is sensitive to the parton distribution functions (PDF) in the proton
in a phase-space region poorly constrained by existing measurements.
It is a benchmark measurement for perturbative quantum chromodynamics phenomenology of heavy quarks and allows constraints to be placed on backgrounds in studies of the Standard Model (SM) Higgs boson and searches
for non-SM physics.

The ATLAS and CMS collaborations reported measurements of Z+b-jet production with jet
transverse momentum\footnote{In this paper we use natural units ($c=\hbar=1$).} larger than $25\gev$ and jet pseudorapidity $|\Peta|<$~2.1, where they 
find good agreement with next-to-leading order (NLO) predictions~\cite{ATLASZb,CMSZb}.
Similar measurements were performed by the CDF~\cite{cdfZb} and D0~\cite{d0Zb} collaborations at the Tevatron,
where the dominant contribution comes from the quark-antiquark interaction. 
The forward acceptance of the LHCb experiment, with a pseudorapidity coverage in the range $2 < \eta < 5$, probes a kinematic region complementary to that probed by ATLAS and CMS. 
The LHCb measurements are sensitive to the parton distribution functions in the proton at low and high values of the Bjorken $x$ variable, where the uncertainties are largest.

In this paper we describe the measurement of the production of Z+b-jet with \decay{\text{Z}/\g^*}{\mup\mun} 
in proton-proton collisions at $\sqs=7\tev$ using the data collected by the LHCb experiment in 2011.
The data set corresponds to an integrated luminosity of 1.0\invfb.

The presence of a bottom hadron candidate is used to tag the jet as originating 
from a bottom quark, following Ref.~\cite{LHCb-PAPER-2014-023}.
The results are compared to NLO and leading-order (LO) calculations using massless and massive bottom quarks.

%% file: detector.tex
\section{Detector and samples}
\label{sec:Detector}

The \lhcb detector~\cite{Alves:2008zz} is a single-arm forward
spectrometer covering the \mbox{pseudorapidity} range $2<\eta <5$,
designed for the study of particles containing \bquark or \cquark
quarks. The detector includes a high-precision tracking system
consisting of a silicon-strip vertex detector surrounding the $pp$
interaction region~\cite{LHCb-DP-2014-001}, a large-area silicon-strip detector located
upstream of a dipole magnet with a bending power of about
$4{\rm\,Tm}$, and three stations of silicon-strip detectors and straw
drift tubes~\cite{LHCb-DP-2013-003} placed downstream of the magnet.
The tracking system provides a measurement of momentum, \ptot,  with
a relative uncertainty that varies from 0.4\,\% at low momentum to 0.6\,\% at 100\gev.
The minimum distance of a track to a primary vertex, the impact parameter, is measured with a resolution of $(15+29/\pt)\mum$,
where \pt is the transverse momentum in \gev. 
Different types of charged hadrons are distinguished using information
from two ring-imaging Cherenkov detectors~\cite{LHCb-DP-2012-003}. Photon, electron and
hadron candidates are identified by a calorimeter system consisting of
scintillating-pad~(SPD) and preshower detectors, an electromagnetic
calorimeter and a hadronic calorimeter. 
The calorimeters have an energy resolution of $\sigma(E)/E=10\%/\sqrt{E}\oplus 1\%$
and $\sigma(E)/E=69\%/\sqrt{E}\oplus 9\%$ (with $E$ in \gev), respectively.
Muons are identified by a
system composed of alternating layers of iron and multiwire
proportional chambers~\cite{LHCb-DP-2012-002}.
The trigger consists of a
hardware stage, based on information from the calorimeter and muon
systems, followed by a software stage, which applies a full event
reconstruction~\cite{LHCb-DP-2012-004}. 

The events used in this analysis are selected by a trigger that requires the presence of at least one muon candidate with $\pt > 10 \gev$. 
In addition, the hardware trigger requires a hit multiplicity in the SPD less than 600, in order to reject events whose processing in the 
software trigger would be too time consuming. This retains about 90\,\% of the events that contain a Z boson.

Simulated samples of $pp$ collisions are generated with 
\pythia v6.4~\cite{Sjostrand:2006za} with a specific LHCb configuration~\cite{LHCb-PROC-2010-056}
using the CTEQ6ll~\cite{CTEQ6ll} parameterization of the PDFs.
Decays of hadronic particles are described by
\evtgen~\cite{Lange:2001uf}, while the interaction of the generated particles with the
detector, and its response, are implemented using 
\geant~\cite{Agostinelli:2002hh} as described in Ref.~\cite{LHCb-PROC-2011-006}.

%% file: strategy.tex
\section{Measurement strategy and event selection}
\label{sec:strategy}

The \decay{\text{Z}}{\mumu} selection follows that described in Ref.~\cite{LHCb-PAPER-2013-058}. 
Muon tracks in the fiducial volume ($2.0 < \Peta(\Pmu) < 4.5$) are required to have transverse
momentum greater than 20\gev. 
In order to have good quality muons, the relative uncertainty on 
the momentum of each muon is required to be less than 10\,\% and 
the \chisq probability for the associated track fit larger than 0.1\,\%. The dimuon candidate mass is required to be in the $60-120$\gev range. 
The contribution from combinatorial background of ($0.31 \pm 0.06$)\,\%, evaluated in Ref.~\cite{LHCb-PAPER-2013-058}, is neglected. 

Charged and neutral particles are clustered by the \antikt algorithm \cite{AntiKT} 
with distance parameter $R=0.5$ as implemented in the \fastjet software package~\cite{FastJet}.
As in Ref.~\cite{LHCb-PAPER-2013-058}, the jet
energy is corrected to the particle level excluding neutrinos and the same jet quality requirements are applied.
The jets are required to be reconstructed within the pseudorapidity range $2.0 < \Peta(\text{jet}) < 4.5$ and 
two transverse momentum thresholds of 10 and 20\gev are studied.
In addition to those kinematic criteria, jets are required to be isolated from the muons of the Z boson decay ($\Delta\text{\it R}(\text{jet},\Pmu)>0.4$),
where $\Delta\text{\it R}$ is the distance in \Peta~--~\Pphi space and \Pphi is the azimuthal angle.

The Z+b-jet cross-section is determined from the ratio of Z+b-jet to Z+jet event yields corrected
for efficiencies and normalized by the Z+jet production cross-section

\begin{equation}
\label{eq:cs}
\sigma(\text{Z+b-jet}) = \frac{\varepsilon({\text{Z+jet}})}{\varepsilon(\text{Z+b-jet})} 
\frac{1}{\varepsilon({\mbox{b-tag}})}
\frac{\text{{\it N}({Z+b-jet}})}{{\it N}(\text{Z+jet})} \sigma(\text{Z+jet}),
\end{equation}

\noindent
where {\it N}(Z+b-jet) is the observed number of Z+b-jet events,
{\it N}(Z+jet) is the number of observed Z+jet events,
$\varepsilon$(Z+jet)/$\varepsilon$(Z+b-jet) is the ratio of efficiencies 
for the reconstruction and selection of Z+jet and Z+b-jet events and $\varepsilon$(b-tag)
is the efficiency of the b-tagging. The production cross-section
of a Z boson associated with jets, $\sigma(\text{Z+jet})$, was previously measured by LHCb~\cite{LHCb-PAPER-2013-058}.
The same data sample, Z boson selection and jet selection are used
 but identification of jets originating from bottom quarks is added. 
By using this approach, the systematic uncertainties and the efficiencies are largely the same as those of Ref.~\cite{LHCb-PAPER-2013-058}, 
except for those related to the b-jet identification.
\begin{figure}[!tb]
  \begin{center}
    \includegraphics[width=0.48\linewidth]{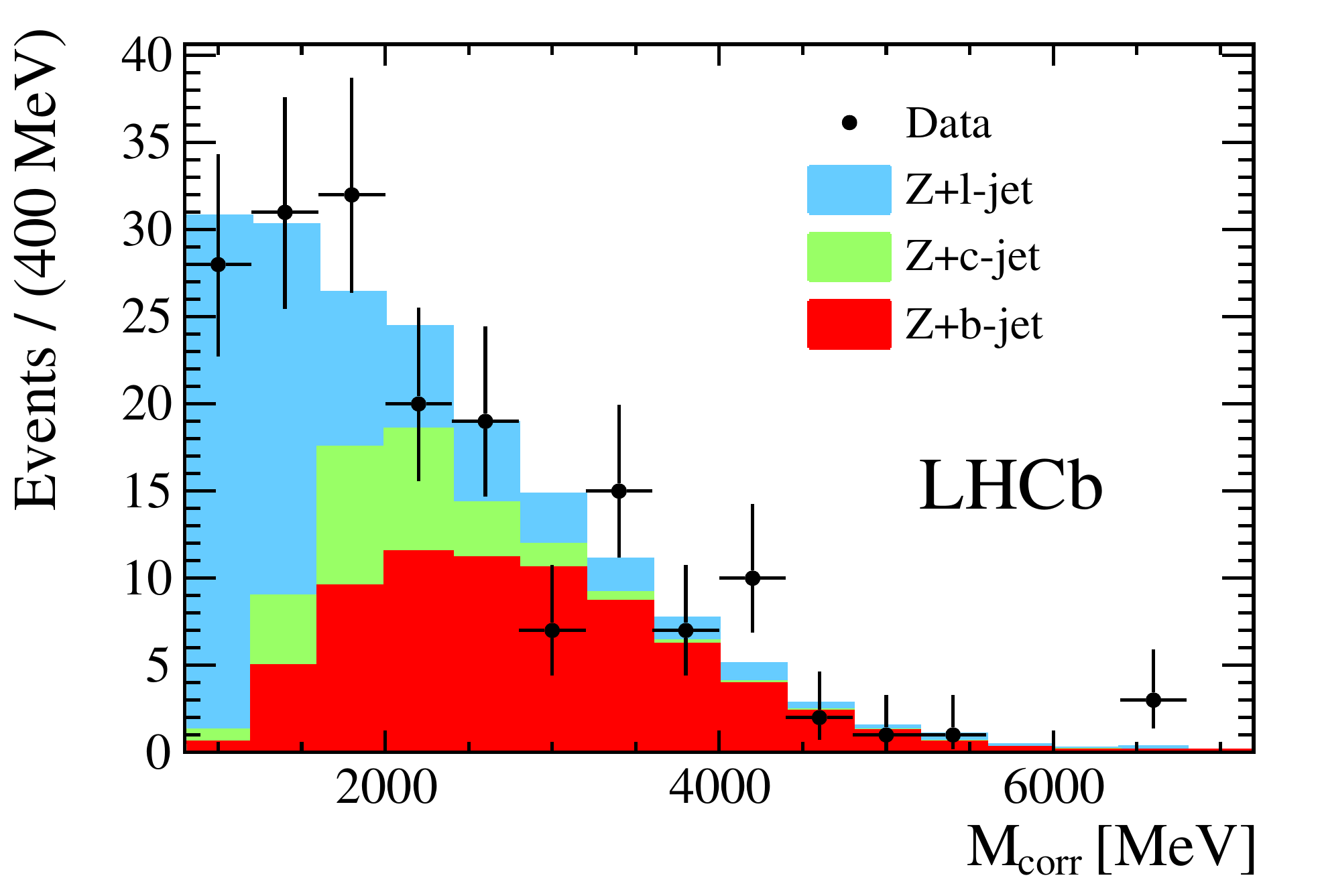}
    \includegraphics[width=0.48\linewidth]{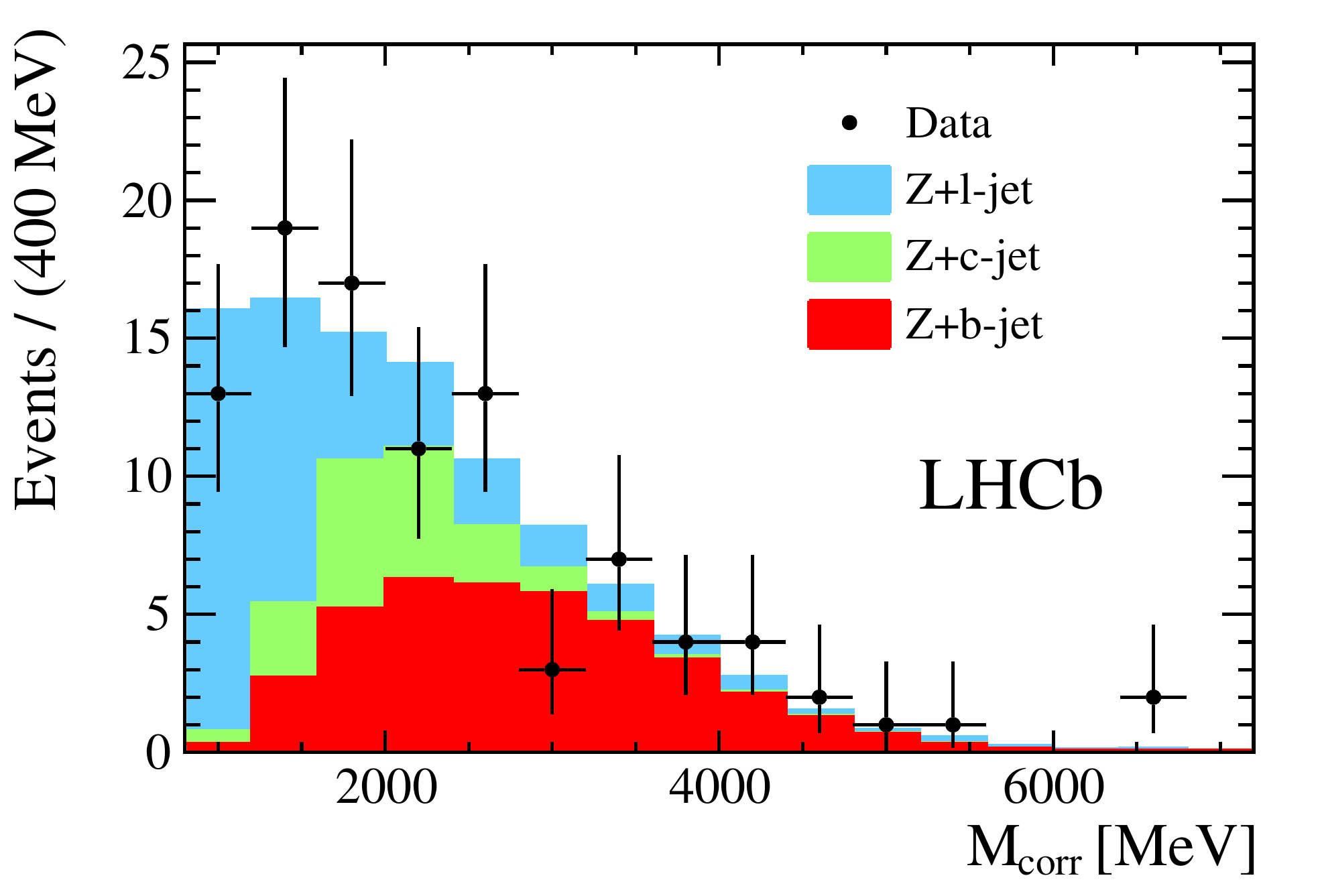}
    \vspace*{-0.5cm}
  \end{center}
  \caption{
      \TopoMCorr distribution for (left) $\pt(\text{jet})>10\gev$ and (right) $\pt(\text{jet})>20\gev$.
      Data (black points) are compared to the template fit results. The uncertainties shown are statistical only.
    }
  \label{fig:fit}
\end{figure}

An algorithm similar to that described in Refs.~\cite{BBDT,LHCb-PAPER-2014-023} is used 
for the identification of secondary vertices consistent with the decay of a 
beauty hadron, using tracks that form the jet. 
Topological secondary vertices (TOPO), significantly
separated from the primary vertex, are formed by considering all combinations
of two, three and four particles within a jet, where particles include both
charged particles reconstruced from tracks and reconstruced $\mathrm{\KS}$ and $\mathrm{\PLambda}$ hadrons.
The requirement of a TOPO candidate greatly reduces the background of
jets originating from light partons (l-jets) and charm quarks (c-jets).
The number of b-jets is extracted from an unbinned likelihood fit to the corrected mass of the TOPO candidate 
defined as ${\TopoMCorr\equiv\sqrt{\TopoM^2+\Pmisssq}+\Pmiss}$.
Here, \TopoM and $p$ are the invariant mass and momentum of the TOPO candidate 
and $\theta$ is the angle between its momentum direction and the flight direction 
inferred from the positions of the primary and secondary vertices~\cite{LHCb-DP-2012-004}. 

Templates for the \TopoMCorr distribution of b-jets, c-jets and l-jets are obtained
from simulation of Z+jet, inclusive b-hadron and inclusive c-hadron production.
The shapes of the templates for b-jets, c-jets and l-jets in these samples 
show no dependence on the production process nor on the \pt  of the jet. 
The \sPlot~method~\cite{Pivk:2004ty} is used to estimate
the b-jet \pt and \Peta spectra.
Figure~\ref{fig:fit} shows the \TopoMCorr distribution 
of b-jet candidates with the fit results overlaid.

Jet reconstruction inefficiencies mainly arise from low-momentum particles
and calorimeter response,
therefore no large differences between jets originating from heavy quarks and from light quarks and gluons are expected. 
Hence, the ratio $\varepsilon(\text{Z+jet})/\varepsilon(\text{Z+b-jet})$ is assumed to be unity,
which is confirmed by simulation.

\begin{figure}[!t]
  \begin{center}
    \includegraphics[width=0.65\linewidth]{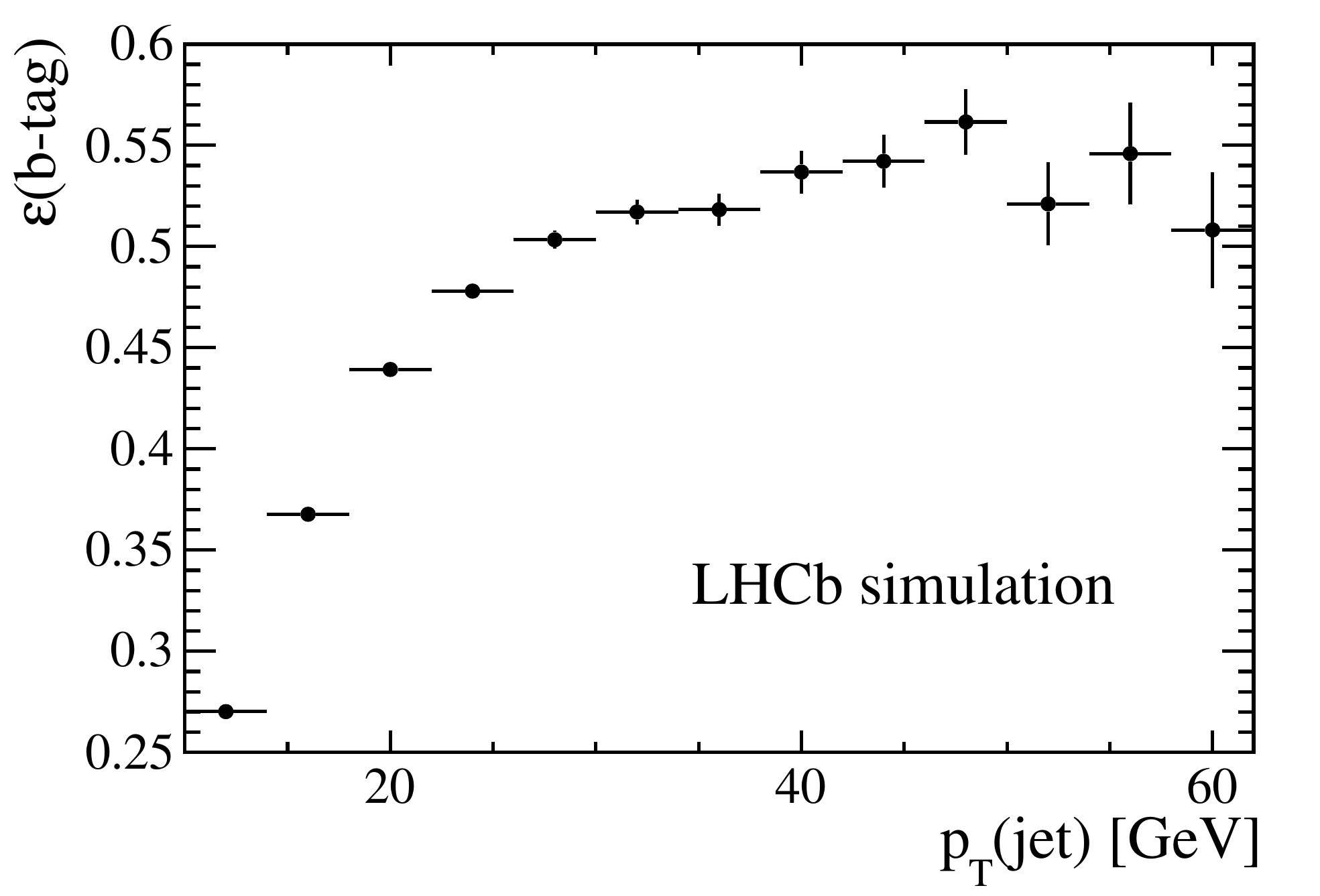}
    \vspace*{-0.5cm}
  \end{center}
  \caption{
    \small 
    Efficiency of b-tagging as function of the jet transverse momentum.
    The uncertainties shown are statistical only.
    }
  \label{fig:beff_eta}
\end{figure}

The b-tagging efficiency, $\varepsilon$(b-tag),
is determined in simulation as a function of the jet transverse
momentum and pseudorapidity. The value of $\varepsilon$(b-tag) shows little variation with pseudorapidity
in the range $2.0 < \Peta(\text{jet}) < 4.5$, while it rises strongly with \pt, reaching a value of 55\,\% at high \pt, 
as shown in Fig.~\ref{fig:beff_eta}. 
The number of Z+b-jet events determined by the template fit is corrected
for the b-tagging efficiency.

%% file: systematics.tex
\section{Systematic uncertainties}
\label{sec:systematics}

The systematic uncertainties related to the Z boson reconstruction, unfolding, jet energy calibration and 
final-state radiation are taken from Ref.~\cite{LHCb-PAPER-2013-058}.
Systematic uncertainties related to the \TopoMCorr templates modelling, b-tagging efficiency and jet 
efficiency flavour dependence are studied in this work.

The systematic uncertainty on the Z boson reconstruction takes account of the contributions from 
the track reconstruction, trigger efficiencies, 
muon identification efficiencies and the model used to fit the Z boson mass. 
The Z boson reconstruction systematic uncertainty is estimated to be 3.5\,\% \cite{LHCb-PAPER-2013-058}.

Migrations in the jet transverse momentum distribution are corrected for by unfolding.
This correction is applied to the value of $\sigma(\text{Z+jet})$ measured in Ref.~\cite{LHCb-PAPER-2013-058} and used in Eq.~\ref{eq:cs}. Detailed studies show that no dedicated unfolding correction is necessary.
The unfolding systematic uncertainty has two contributions. 
The difference between the SVD\cite{paper:SVD} and D'Agostini\cite{paper:D'Agostini} 
unfolding methods is assigned as one contribution. 
The second contribution comes from the difference between the unfolded distribution 
and the true distribution in an independent simulated sample. 
This systematic uncertainty is taken from Ref.~\cite{LHCb-PAPER-2013-058} and it is evaluated to be 1.5\,\%.

An important contribution to the systematic uncertainties related to the 
jets comes from the jet-energy scale. It is estimated by comparing 
the transverse momentum of the Z boson and the jet in single jet events, 
where their momenta are azimuthally opposed, and are expected to be balanced.  
An additional contribution of 2\,\% to the jet-energy scale uncertainty comes 
from the differences between jets initiated from quarks and gluons. 
The systematic uncertainty of the jet identification is estimated by
comparing the number of candidates in data and simulation with a more stringent selection. 
The total systematic uncertainty related to jets is 7.8\,\% as estimated in Ref.~\cite{LHCb-PAPER-2013-058}.

The systematic uncertainty associated to final-state radiation is obtained
by direct comparison to the simulation described above and an additional simulation, using HERWIG++\cite{paper:HERWIG}, as in Ref.~\cite{LHCb-PAPER-2013-058}; it is estimated to be 0.2\,\%. 
The systematic uncertainty associated with the knowledge of the luminosity is estimated to be 3.5\,\%~\cite{LHCb-PAPER-2011-015}. 

Possible systematic variations of the final result due
to the extraction of $\varepsilon$(b-tag) and \TopoMCorr templates
from simulations are controlled using two data samples enriched in b-jets and c-jets.
The b-jet (c-jet) enriched sample is selected via 
one \Bpm (\Dpm) hadron candidate decaying to $\jpsi\Kpm$ ($\Kmp\pipm\pipm$)
produced with a large azimuthal opening angle with respect to a probe jet.
The b-tagging requirement is applied to the probe jet and a template fit is performed. 
Three studies are performed: 
1) the data are divided into two ranges of \TopoM, the template fit is performed on each and the sum of the 
resulting b-jet yields is compared with the standard result; 
2) a looser b-tagging requirement is applied
and the b-jet yields after b-tagging efficiency correction are compared with the default values; and 
3) the \TopoMCorr template is smeared to account for possible differences between simulation and data,
and the impact on the b-jet yields is studied.
The \TopoMCorr simulation modelling and
TOPO candidate reconstruction efficiency studies are found to affect this measurement by up to 15\,\%,
where this uncertainty is dominated by the first of the studies mentioned above.

Using simulation, $\varepsilon(\text{Z+jet})/\varepsilon(\text{Z+b-jet})$ 
is found to be compatible with unity within 2\,\%, which is taken as the systematic uncertainty 
due to the flavour dependence of the jet efficiency.

The systematic uncertainties are summarized in Table \ref{tab:sys}. They are added in quadrature
leading to a total systematic error of 17.8\,\%.

\begin{table} [tb]
\caption{\small
Relative systematic uncertainty considered for the Z+b-jet cross-section for $\pt($jet) $>$ 20\gev. 
The relative uncertainties are similar for the 10\gev threshold. The first four contributions are from Ref.~\cite{LHCb-PAPER-2013-058}.
}
\label{tab:sys}
\begin{center}
\begin{tabular}{l r}
Source of systematic uncertainty  & Relative uncertainty (\%)  \\
\hline
Z boson reconstruction & 3.5 \\
Unfolding & 1.5 \\
Jet-energy scale, resolution and reconstruction & 7.8 \\
Final-state radiation & 0.2 \\
Luminosity &￼3.5 \\
\TopoMCorr template and b-tagging & 15.0 \\
Jet reconstruction flavour dependence & 2.0 \\
\hline
Total &￼17.8 \\
\end{tabular}
\end{center}
\end{table}

%% file: results.tex
\section{Results}

We observe 179 (97) Z+jet events where at least one jet 
fulfils the b-tagging requirement for the $\pt\text{(jet)}>10\gev$ ($20\gev$) threshold.
No events with more than one b-tagged jet are observed.
The extended unbinned likelihood fit of the \TopoMCorr spectrum using Z+l-jet,
Z+c-jet and Z+b-jet templates determines $72\pm15$ 
($39\pm11$) Z+b-jet events for
the $\pt\text{(jet)}>10\gev$ ($20\gev$) threshold.
The number of candidates corrected for b-tagging efficiency
is found to be 
$177\pm36$ ($76\pm21$)
for the $\pt\text{(jet)}>10\gev$ ($20\gev$) threshold.
Using the measurements of Ref.~\cite{LHCb-PAPER-2013-058}, we determine
the cross-section of Z+b-jet production to be
\begin{equation*}
  \sigma(\text{$\text{Z}/\g^*(\mup\mun)$+b-jet}) = 295 \pm 60~(\text{stat}) \pm 51~(\text{syst}) \pm 10~(\text{lumi}) \fb
\end{equation*}
for $\pt($jet) $>$ 10\gev, and  
\begin{equation*}
  \sigma(\text{$\text{Z}/\g^*(\mup\mun)$+b-jet}) =  128 \pm 36~(\text{stat}) \pm 22~(\text{syst}) \pm 5~(\text{lumi}) \fb
\end{equation*}
for $\pt($jet) $>$ 20\gev.  These cross-sections are evaluated within the fiducial 
region $\pt(\mu) > 20\gev$, $60 \gev < M(\mun\mup) < 120 \gev$, $2.0 < \eta(\text{jet}) < 4.5$, $2.0 < \eta(\mu) < 4.5$ and $\Delta R(\text{jet}, \mu) > 0.4$.

\begin{figure}[htb]
  \begin{center}
    \includegraphics[width=0.8\linewidth]{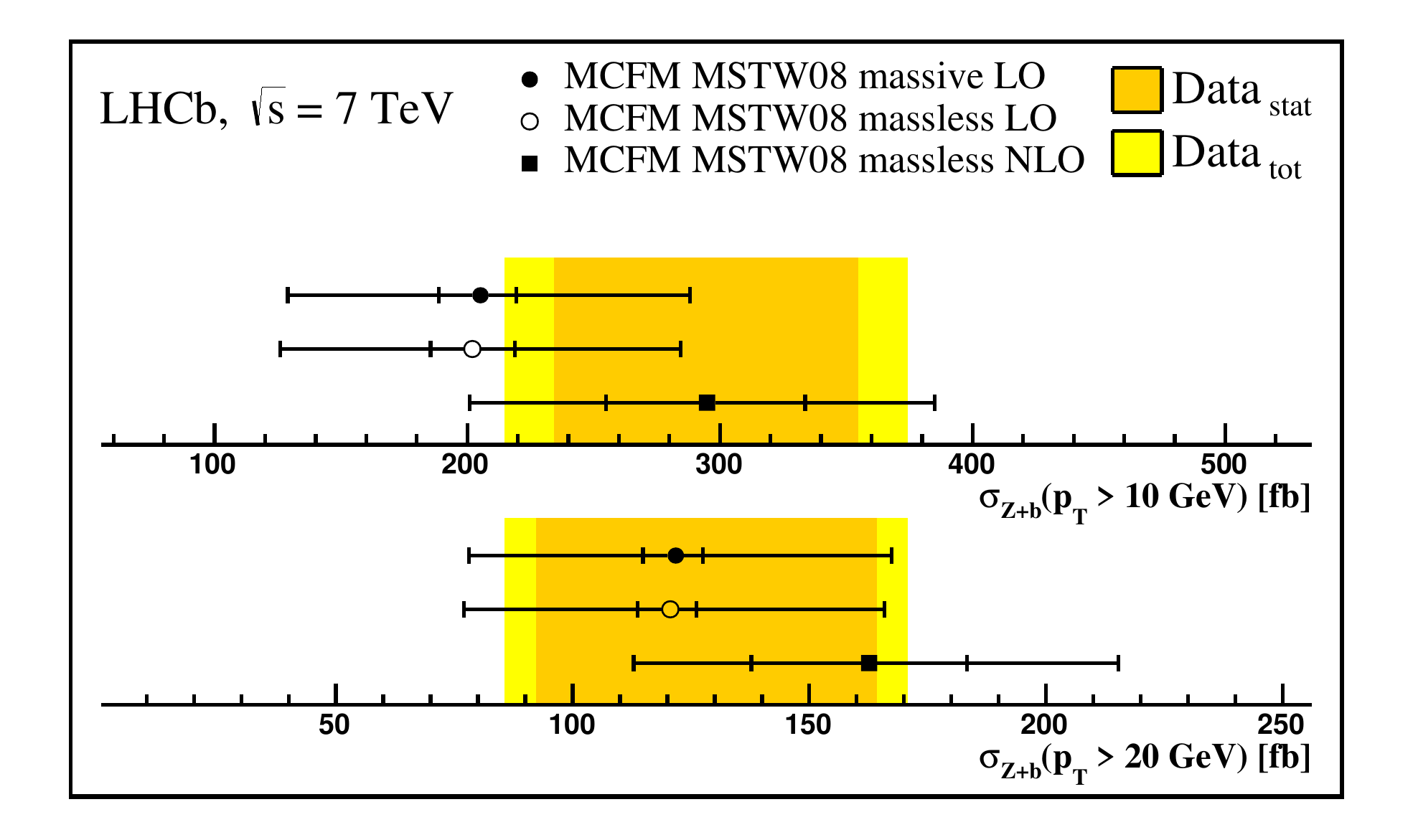}
    \vspace*{-0.5cm}
  \end{center}
  \caption{
Z+b-jet cross-section for two $\pt(\text{jet})$ thresholds.
The colour band shows the LHCb measurement (with the inner orange band 
showing the statistical uncertainty, and the outer yellow band showing the total uncertainty). 
The points with error bars correspond to the theoretical predictions with the inner error bars indicating their PDF uncertainties. 
These cross-sections are evaluated within the fiducial region $\pt(\mu) > 20\gev$, $60 \gev < M(\mun\mup) < 120 \gev$, $2 < \eta(\text{jet}) < 4.5$, $2 < \eta(\mu) < 4.5$ and $\Delta R(\text{jet}, \mu) > 0.4$.
 }
  \label{fig:zbcs}
\end{figure}

The measurements are compared to predictions of the Z+b-jet cross-section calculated 
using \mcfm~\cite{MCFM} in the same kinematic range as for this measurement.
The uncertainties include the PDF and theory uncertainties evaluated 
by varying independently the renormalization and factorization scales by a factor two around their nominal scales. 
Neither showering nor hadronization are included in MCFM; therefore the same kinematic requirements applied
to jets in the data analysis are applied to the bottom quarks in \mcfm.
An overall correction is calculated by generating Z+b-jet events with \pythia v8.170
with the MSTW08 \PDF set~\cite{MSTW08} where the same acceptance requirements are applied. 
Jets are reconstructed with \fastjet using the \antikt algorithm with $R=0.5$
and then matched with a bottom quark, requiring $\Delta R\text{(jet,b-quark)}<0.5$. The ratio between the number of events with 
at least one b-jet that fulfils the kinematic requirements of this measurement
and the number of events with at least one b quark within the 
acceptance criteria are used as the fragmentation and hadronization correction for the MCFM predictions.
The ratio is 0.77 (0.90) for $\pt\text{(jet)} > 10~(20) \gev$.
Figure ~\ref{fig:zbcs} shows the cross-section measurements compared to 
the LO calculation with massive bottom quarks and to LO and NLO calculations neglecting the bottom quark mass.

%% file: conclusion.tex
\section{Summary}

The cross-section for forward production of a Z boson or an off-shell photon, in the $\mup\mun$ channel, and a bottom-quark is measured in $\sqrt{s}=7\tev$ proton-proton collisions corresponding to an integrated luminosity of 1.0\invfb of data collected in 2011
by the LHCb collaboration.
Results are reported for the kinematic region
$2.0 < \Peta(\Pmu) < 4.5$, $\pt(\Pmu) > 20 \gev$, $60 < M(\mup\mun) < 120 \gev$,
$\pt(\text{jet})>10(20)\gev$, $2.0 < \Peta(\text{jet}) < 4.5$ and $\Delta\text{\it R}(\text{jet},\Pmu)>0.4$. 
The measured cross-sections are
\begin{equation*}
  \sigma(\text{$\text{Z}/\g^*(\mup\mun)$+b-jet}) = 295 \pm 60~(\text{stat}) \pm 51~(\text{syst}) \pm 10~(\text{lumi}) \fb
\end{equation*}
for $\pt$(jet)$>10\gev$, and
\begin{equation*}
  \sigma(\text{$\text{Z}/\g^*(\mup\mun)$+b-jet}) =  128 \pm 36~(\text{stat}) \pm 22~(\text{syst}) \pm 5~(\text{lumi}) \fb
\end{equation*}
for $\pt$(jet)$>20\gev$.

The results are in agreement with MCFM predictions for massless and massive bottom quark
calculations.

\FloatBarrier

%% file: acknowledgements.tex
\section*{Acknowledgements}

%
 
\noindent We express our gratitude to our colleagues in the CERN
accelerator departments for the excellent performance of the LHC. We
thank the technical and administrative staff at the LHCb
institutes. We acknowledge support from CERN and from the national
agencies: CAPES, CNPq, FAPERJ and FINEP (Brazil); NSFC (China);
CNRS/IN2P3 (France); BMBF, DFG, HGF and MPG (Germany); SFI (Ireland); INFN (Italy); 
FOM and NWO (The Netherlands); MNiSW and NCN (Poland); MEN/IFA (Romania); 
MinES and FANO (Russia); MinECo (Spain); SNSF and SER (Switzerland); 
NASU (Ukraine); STFC (United Kingdom); NSF (USA).
The Tier1 computing centres are supported by IN2P3 (France), KIT and BMBF 
(Germany), INFN (Italy), NWO and SURF (The Netherlands), PIC (Spain), GridPP 
(United Kingdom).
We are indebted to the communities behind the multiple open 
source software packages on which we depend. We are also thankful for the 
computing resources and the access to software R\&D tools provided by Yandex LLC (Russia).
Individual groups or members have received support from 
EPLANET, Marie Sk\l{}odowska-Curie Actions and ERC (European Union), 
Conseil g\'{e}n\'{e}ral de Haute-Savoie, Labex ENIGMASS and OCEVU, 
R\'{e}gion Auvergne (France), RFBR (Russia), XuntaGal and GENCAT (Spain), Royal Society and Royal
Commission for the Exhibition of 1851 (United Kingdom).

%% file: LHCb_authorlist.tex
\centerline{\large\bf LHCb collaboration}
\begin{flushleft}
\small
R.~Aaij$^{41}$, 
B.~Adeva$^{37}$, 
M.~Adinolfi$^{46}$, 
A.~Affolder$^{52}$, 
Z.~Ajaltouni$^{5}$, 
S.~Akar$^{6}$, 
J.~Albrecht$^{9}$, 
F.~Alessio$^{38}$, 
M.~Alexander$^{51}$, 
S.~Ali$^{41}$
% ...

{\footnotesize \it
% ...
$ ^{5}$Clermont Universit\'{e}, Universit\'{e} Blaise Pascal, CNRS/IN2P3, LPC, Clermont-Ferrand, France\\
$ ^{6}$CPPM, Aix-Marseille Universit\'{e}, CNRS/IN2P3, Marseille, France\\
% ...
$ ^{9}$Fakult\"{a}t Physik, Technische Universit\"{a}t Dortmund, Dortmund, Germany\\
% ...
$ ^{37}$Universidad de Santiago de Compostela, Santiago de Compostela, Spain\\
$ ^{38}$European Organization for Nuclear Research (CERN), Geneva, Switzerland\\
% ...
$ ^{41}$Nikhef National Institute for Subatomic Physics, Amsterdam, The Netherlands\\
% ...
$ ^{46}$H.H. Wills Physics Laboratory, University of Bristol, Bristol, United Kingdom\\
% ...
$ ^{51}$School of Physics and Astronomy, University of Glasgow, Glasgow, United Kingdom\\
$ ^{52}$Oliver Lodge Laboratory, University of Liverpool, Liverpool, United Kingdom\\
% ...
}
\end{flushleft}